\documentstyle[aps,prl,twocolumn]{revtex}
\begin{document}
\draft
\title{Recovery temperature for nonclassical energy transfer in
atom-surface scattering}
\author{B. Gumhalter$^{a,b}$, A. \v{S}iber$^{b}$ and J.P. Toennies$^{c}$}
\address{ $^a$ Abdus Salam International Centre for Theoretical Physics, 
Trieste, Italy,\protect\\
$^b$ Institute of Physics of the University, P.O. Box 304, 10001
Zagreb, Croatia, \protect\\
$^c$ Max-Planck-Institut f\"{u}r Str\"{o}mungsforschung, D-37073 
G\"{o}ttingen,
Germany}
 
\maketitle

\begin{abstract}
Nonperturbative expressions are derived for the angular resolved energy
transfer spectra in the quantum regime of multiphonon scattering of
inert gas atoms 
from surfaces. Application to  He atom scattering from a 
prototype heatbath Xe/Cu(111) shows good agreement with experiments.
This enables a full quantum calculation of the total energy transfer 
$\mu$ and therefrom the much debated recovery or equilibrium 
temperature $T_{r}$ characteristic of zero energy transfer in gas-surface collisions in the free molecular flow regime. Classical 
universal character of $\mu$ and $T_{r}$ is refuted.
\end{abstract}

\pacs{PACS numbers: 68.35.Ja, 34.50.Dy, 63.22.+m, 47.45.Nd}


Despite the great progress recently made in understanding
the dynamics of gas-surface collisions \cite{Hulpke} little effort has 
gone into applying this knowledge to problems of general relevance 
\cite{Legge}. Because of their fundamental importance for 
a wide range of the flow phenomena \cite{Bird} there is a need to be
able to predict the magnitudes of the heat transfer $\mu$ \cite{Schaaf}
and the recovery temperature $T_{r}$ at which zero energy
transfer to the surface occurs in the free molecular flow regime
\cite{Cerc}, but the classically calculated $\mu$ and $T_{r}$ generally
fail to reproduce the experimental data \cite{Legge}. However, the 
information on single-  and multi-phonon 
excitations recently accumulated from He atom scattering (HAS) from
surfaces \cite{Hofmann,Sibener,GrahamJCP,BraunPRL,Xe} combined with
novel theoretical 
developments now opens up the possibility of calculating the heat transfer 
on a microscopic level and in the fully quantum scattering regime. In
this Letter a new approach based on multiphonon scattering theory 
\cite{BGL,HAS,comment} is developed to
predict the energy
transfer in interactions of He atom beams with Xe monolayers on Cu(111)
which represent an ideal prototype heatbath encompassing both the
dispersive and nondispersive surface phonons \cite{Xe}. 
The calculations reproduce the HAS data remarkably well and thus enable
a reliable prediction of the recovery temperature which is
found to differ markedly from the results of classical accommodation 
theories currently in use \cite{Schaaf,Cerc}.

The total energy transfer $\mu$ which enters the heat transfer and 
accommodation coefficients \cite{Legge,Schaaf} is evaluated from 
\begin{equation} 
\mu({\bf k_{i}},T_{s})=\int^{\infty}_{-\infty}\varepsilon N_{{\bf k_{i}},
T_{s}}(\varepsilon) d\varepsilon.
\label{eq:mu1}
\end{equation}
where $N_{{\bf k_{i}},T_{s}}(\varepsilon)$ is the scattering spectrum
which gives 
the probability density for an atom with initial momentum 
$\hbar{\bf k_{i}}$ to exchange energy $\varepsilon$ with  
a surface at the temperature $T_{s}$. The final atom state can be either a continuum $\mid c\rangle$ or a bound state $\mid b\rangle$ of the static atom-surface potential $U({\bf r})$\cite{BB,Boeheim,commscat}.
However, there are two major difficulties in applying Eq. 
(\ref{eq:mu1}) to atom-surface scattering. First,   
$N_{{\bf k_{i}},T_{s}}(\varepsilon)$ and thereby $\mu({\bf k_{i}},T_{s})$ 
are not directly accessible in typical HAS time-of-flight (TOF) 
measurements from which most of the data are available at present. 
These experiments yield the {\em energy and angular resolved}
quantities usually only for fixed total 
scattering angle $\theta_{SD}=\theta_{i}+\theta_{f}$, where
$\theta_{i}$ ($\theta_{f}$) is the initial (final) scattering angle.
The TOF spectrum
is directly proportional to the $c\rightarrow c$ component of the full energy and 
parallel momentum  resolved scattering distribution 
$N_{{\bf k_{i}},T_{s}}(\varepsilon,\Delta{\bf K})$
\cite{comment} which in turn is related to the 
{\em angular resolved} energy transfer 
\begin{equation}
\mu_{r}({\bf k_{i}},T_{s},\theta_{f})=
\frac{\int \varepsilon N^{c\rightarrow c}_{{\bf k_{i}},T_{s}}(\varepsilon,\Delta{\bf
K}(\varepsilon))d\varepsilon}
{\int  N^{c\rightarrow c}_{{\bf k_{i}},T_{s}}(\varepsilon,\Delta{\bf K}(\varepsilon))
d\varepsilon}.
\label{eq:muresol}
\end{equation}
As $\mu_{r}({\bf k_{i}},T_{s},\theta_{f})$ can be computed from both the theoretical $N^{c\rightarrow c}_{{\bf k_{i}},T_{s}}(\varepsilon,\Delta{\bf K})$ and
the experimental TOF-spectra a direct comparison
of the two results enables the verification of model calculations. The 
second difficulty arises in  calculations of the reliable multiphonon 
$N_{{\bf k_{i}},T_{s}}(\varepsilon)$, obtained from 
$N_{{\bf k_{i}},T_{s}}(\varepsilon,\Delta{\bf K})$ by integration over 
$\Delta{\bf K}$, in the regime in which it is important to
treat the dynamics 
of both the projectile and surface vibrations quantum mechanically.  
However, the recent progress in interpreting the multiphonon HAS from 
monolayers of noble gas atoms (Ar, Kr, Xe) adsorbed on metals 
\cite{Sibener,GrahamJCP,BraunPRL,Xe} makes it now possible to exactly 
calculate $N_{{\bf k_{i}},T_{s}}(\varepsilon,\Delta{\bf K})$ and
thereby assess the total energy transfer in the studied collisions. 
These adlayers sustain low-energy longitudinal (L) and
shear-horizontal (SH) in-plane polarized modes with acoustic-like
dispersion and
nondispersive vertically polarized 
Einstein-like modes (S). 
The presence of nondispersive modes gives rise to special
interference effects in $N_{{\bf k_{i}},T_{s}}(\varepsilon,\Delta{\bf K})$
which requires its calculation in a closed, nonperturbative form.  Such
a solution is presented below and applied to the study of energy
transfer in HAS 
from $(\sqrt{3}\times\sqrt{3})R30^{o}$ monolayers of 
Xe on Cu(111) which exhibit very weak diffraction and hence can be 
considered as flat in energy exchange processes \cite{Xe}. This suppresses selective adsorption assisted energy transfer\cite{Boeheim} and greatly
simplifies calculations and comparisons of  
$N^{c\rightarrow c}_{{\bf k_{i}},T_{s}}(\varepsilon,\Delta{\bf K})$ with the TOF spectra.

In the scattering regime in which uncorrelated phonon processes are 
dominant the angular resolved scattering spectrum has the following unitary 
form  
\cite{BGL,HAS}:
\begin{eqnarray}
N_{{\bf k_{i}},T_{s}}(\varepsilon,\Delta{\bf K})
&=& 
\int^{\infty}_{-\infty} \frac{d\tau d^{2}{\bf
R}}{(2\pi\hbar)^{3}}
e^{i[\varepsilon\tau-\hbar(\Delta{\bf K}) {\bf R}]/\hbar}\nonumber\\
&\times&
\exp[2W(\tau,{\bf R})-2W].
\label{eq:specEBA}
\end{eqnarray}
Here $2W(\tau,{\bf R})$ is the scattering function\cite{HAS}, 
$2W=2W(\tau=0,{\bf R}=0)$ 
is the exponent of the Debye-Waller factor (DWF) \cite{dwf} and 
$\tau$ and 
${\bf R}$ are auxiliary integration variables used to
project the states with $\varepsilon$ and $\Delta{\bf K}$ from the integral
on the RHS of Eq. (\ref{eq:specEBA}). 
In the range of validity of Eq. (\ref{eq:specEBA}) the energy and parallel 
momentum are conserved in each phonon exchange process and 
$\varepsilon$ and $\Delta{\bf K}$ are constrained to satisfy the total
energy and parallel momentum conservation in the collision.  
Using expression (\ref{eq:specEBA}) it is possible to write Eq. 
(\ref{eq:mu1}) as:
\begin{equation}
\mu({\bf k_{i}},T_{s})=i\frac{\partial}{\partial \tau} 
2W(\tau=0,{\bf R}=0),
\label{eq:mu1EBA}
\end{equation}
which can be readily calculated once $2W(\tau,{\bf R})$ is established.
In the following the projectile-phonon coupling is assumed linear in
the adsorbate displacements since this gives the
dominant multiphonon contribution observed in HAS \cite{AM}. This
yields \cite{HAS}:
\begin{eqnarray}
2W(\tau,{\bf R})&=&
\sum_{{\bf Q,G},j,k_{z}}
\left[ \mid {\cal V}^{{\bf K_{i},Q+G}}_{k_{z},k_{zi},j}(+)\mid^{2}
[\bar{n}(\omega_{{\bf Q}j})+1]\right.\nonumber\\
&\times&  e^{-i[\omega_{{\bf Q}j}\tau -
{\bf (Q+G)R}]}
\nonumber\\ 
&+&
\left. \mid {\cal V}^{{\bf K_{i},Q+G}}_{k_{z},k_{zi},j}(-) \mid^{2}\bar{n}
(\omega_{{\bf Q}j})  e^{i[\omega_{{\bf Q}j}\tau -{\bf (Q+G)R}]}\right],
\label{eq:WEBA}
\end{eqnarray}
where ${\bf Q}$, $j$  and $\omega_{{\bf Q}j}$ denote the parallel 
wave-vector, branch index and frequency of a 
normal phonon mode, respectively, and ${\bf G}$ is the adlayer
reciprocal lattice vector.  
${\bf k}=({\bf K},k_{z})$ where $k_{z}$ is the quantum number describing projectile perpendicular motion in $c$- or $b$-states of $U({\bf r})$.   $\bar{n}(\omega_{{\bf Q}j})$ 
is the Bose-Einstein distribution,
and ${\cal V}^{{\bf
K_{i},Q+G}}_{k_{z},k_{zi},j}(\pm)$ denote 
one-phonon emission $(+)$ and absorption $(-)$  on-shell scattering
matrix elements normalized to particle current in
the $z$-direction
\cite{Xe,HAS}. 
Substitution of expression (\ref{eq:WEBA}) into (\ref{eq:mu1EBA}) gives 
\begin{equation}
\mu({\bf k_{i}},T_{s})=\mu_{0}({\bf k_{i}})+\mu_{rec}({\bf k_{i}},T_{s}),
\label{eq:musum}
\end{equation}
in which the temperature independent part is 
\begin{equation}
\mu_{0}({\bf k_{i}})=
\sum_{{\bf Q,G},j,k_{z}}\hbar\omega_{{\bf Q}j}
\mid {\cal V}^{{\bf K_{i},Q+G}}_{k_{z},k_{zi},j}(+) 
\mid^{2},
\label{eq:mu0}
\end{equation}
and the $T_{s}$-dependence is determined by the recoil term 
\begin{eqnarray}
\mu_{rec}({\bf k_{i}},T_{s})
&=&
\sum_{{\bf Q,G},j,k_{z}}\hbar\omega_{{\bf Q}j}
\left[ \mid {\cal V}^{{\bf K_{i},Q+G}}_{k_{z},k_{zi},j}(+) 
\mid^{2}\right.
\nonumber\\
&-&\left.
 \mid {\cal V}^{{\bf K_{i},Q+G}}_{k_{z},k_{zi},j}(-) \mid^{2}\right]
\bar{n}(\omega_{{\bf Q}j}),
\label{eq:murec}
\end{eqnarray}
which vanishes in the recoilless trajectory approximation (TA) for the 
projectile motion \cite{CelliHimes}. However, the TA
may fail even for heavier atoms \cite{Burke,commscat} and in the present 
quantum scattering regime the recoil is large (c.f. Figs. 10 and 11 in Ref. 
\cite{Xe}) making $\mu({\bf k_{i}},T_{s})$ strongly $T_{s}$-dependent. 
 
\vskip 7 cm

\begin{figure}
\caption{Comparison of the 
temperature dependence of the {\em angular resolved}  energy transfer 
$\mu_{r}({\bf k_{i}},T_{s},\theta_{f})$ calculated
from He$\rightarrow$Xe/Cu(111) TOF spectra for four experimental
$E_{i}$ and fixed scattering
geometry (open symbols), and from the present model (full lines). The inset
shows a comparison of the measured and calculated multiphonon scattering
spectrum for 
$E_{i}=45.11$ meV, $\theta_{i}=50^{0}$, $T_{s}=58.2$ K.} 
\label{recovfg1}
\end{figure}

The nonperturbative multiphonon solution for
$N_{{\bf k_{i}},T_{s}}(\varepsilon,\Delta{\bf K})$, 
which enables explicit calculation of expression 
(\ref{eq:muresol}) and thus a direct comparison with the HAS 
data, is obtained by separating the most strongly coupling
Einstein branch $j=S$ of frequency $\omega_{S}$ 
out of the sum in Eq. (\ref{eq:WEBA}), giving: 
\begin{equation}
\exp[2W(\tau,{\bf R})-2W]=
N^{Ein}_{{\bf k_{i}},T_{s}}(\tau,{\bf R})
N^{dis}_{{\bf k_{i}},T_{s}}(\tau,{\bf R}),
\label{eq:NN}
\end{equation}
where $dis$ denotes the remaining 
dispersive modes.
Employing trigonometric identities to $N^{Ein}_{{\bf
k_{i}},T_{s}}(\tau,{\bf R})$ yields:
\begin{equation}
N^{Ein}_{{\bf k_{i}},T_{s}}(\tau,{\bf R})=
e^{-2W_{S}}\sum_{l=-\infty}^{\infty}P_{l}({\bf R})e^{-il\omega_{S}\tau},
\label{eq:Ntau}
\end{equation}
where $2W_{S}=2W_{S}(\tau=0,{\bf R=0})$ is the corresponding
Debye-Waller exponent and
\begin{eqnarray}
P_{l}({\bf R})
&=&
\left[\left(\sqrt{ \frac{[\bar{n}(\omega_{S})+1]
{\cal V}_{S}^{2}({\bf
R},+)}{\bar{n}(\omega_{S}){\cal V}_{S}^{2}({\bf R},-)}}\right)^{l}\right.
\nonumber\\
&\times&
\left. I_{l}\left(\sqrt{4\bar{n}(\omega_{S})[\bar{n}(\omega_{S})+1]{\cal V}_{S}^{2}
({\bf R},+){\cal V}_{S}^{2}({\bf R},-)}\right)\right],
\label{eq:Pl}
\end{eqnarray}
where ${\cal V}_{S}^{2}
({\bf R},\pm)=\sum_{{\bf Q,G},k_{z}}
\mid {\cal V}^{{\bf K_{i},Q+G}}_{k_{z},k_{zi},S}(\pm)\mid^{2} 
e^{\mp i[\omega_{S}\tau -{\bf (Q+G)R}]}$ 
and $I_{l}$ is the modified Bessel function of the first kind.
Hence, the separated Einstein phonon component of the spectrum in Eq.
(\ref{eq:specEBA})  takes the form
\begin{equation}
N_{{\bf k_{i}},T_{s}}^{Ein}(\varepsilon,\Delta{\bf K})=e^{-2W_{S}}
\sum_{l=-\infty}^{\infty}N_{l}(\Delta{\bf
K})\delta(\varepsilon -l\hbar\omega_{S}),
\label{eq:NEinst}
\end{equation}
where $N_{l}(\Delta{\bf K})=\int d^{2}{\bf R} e^{-i(\Delta{\bf K}){\bf
R}} P_{l}({\bf R})/(2\pi)^{2}$ is the intensity of the $l$-th Einstein 
multiphonon loss 
($l>0$) or gain ($l<0$) peak. The elastic intensity 
$N_{{\bf k_{i}},T_{s}}^{Ein}(\varepsilon=0,\Delta{\bf K}\neq0)$ is 
non-vanishing for
 $T_{s}>0$ because multiple exchange of nondispersive phonons may give 
rise to 
finite momentum transfer without a net energy transfer. 
However, the spectral intensity of
 the specular elastic peak is
$e^{-2W_{S}}\delta(\Delta{\bf
K})$ because $P_{0}({\bf R\rightarrow\infty})\rightarrow 1$. 
On the other hand, the elastic peak in $N_{{\bf
k_{i}},T_{s}}(\varepsilon)$, obtained from Eq. (\ref{eq:NEinst}) by 
integrating over all $\Delta{\bf K}$, is weighed by 
$e^{-2W_{S}}P_{0}({\bf R=0})$ but with $P_{0}({\bf R=0})\neq 1$ due to 
the same multiquantum exchange effect.

A similar procedure yields for $N^{dis}_{{\bf k_{i}},T_{s}}(\tau,{\bf
R})$ in Eq.  (\ref{eq:NN}):
\begin{eqnarray}
N^{dis}_{{\bf k_{i}},T_{s}}(\tau,{\bf R})
 & = &
e^{-2W_{dis}}
\exp\left[\sum_{{\bf Q,G},k_{z}',j\neq S} 
\sqrt{4[\bar{n}(\omega_{{\bf Q}j})+1]
\bar{n}(\omega_{{\bf Q}j})
\mid{\cal V}^{{\bf K_{i},Q+G}}_{k_{z}',k_{zi},j}(+)
\mid^{2}\mid{\cal V}^{{\bf K_{i},Q+G}}_{k_{z}',k_{zi},j}
(-)\mid^{2}} \right. 
\nonumber\\
 &\times&  
\left. \cos\left(i\ln\sqrt{\frac{[\bar{n}(\omega_{{\bf Q}j})+1]
\mid{\cal V}^{{\bf K_{i},Q+G}}_{k_{z}',k_{zi},j}(+)
\mid^{2}}{\bar{n}(\omega_{{\bf Q}j})
\mid {\cal V}^{{\bf K_{i},Q+G}}_{k_{z}',k_{zi},j}(-)\mid^{2}}}
+[\omega_{{\bf Q}j}\tau-{\bf (Q+G)R}]\right)\right]. 
\label{eq:Nj}
\end{eqnarray}
Expressions  (\ref{eq:Ntau})-(\ref{eq:Nj}) are 
{\em exact} within the validity of (\ref{eq:specEBA}) \cite{HAS} and
include the {\em combined effect of recoil and temperature} on
the multiphonon scattering spectra. 

The calculations of expressions (\ref{eq:WEBA})-(\ref{eq:Nj}) were
carried out by modeling $U({\bf r})=U(z)$, the projectile-phonon coupling and
the dynamical matrix of the Xe monolayer and 40 layer thick Cu(111) 
slab as in Refs. 
\cite{BraunPRL,Xe}.  
The calculated 
$N^{c\rightarrow c}_{{\bf k_{i}},T_{s}}(\varepsilon,\Delta{\bf K})$ and  
$\mu_{r}({\bf k_{i}},T_{s},\theta_{f})$ in the multiphonon scattering 
regime were tested by directly comparing with experiment. 
Figure \ref{recovfg1} shows a comparison of the angular resolved energy 
transfer obtained from Eq. (\ref{eq:muresol}) without
invoking any
fit parameters, with the values calculated by integrating the TOF
spectra for fixed $\theta_{SD}$
\cite{BraunPRL,Xe}. The inset shows a comparison of the
experimental and calculated scattering spectrum (\ref{eq:specEBA}) for
one particular set of the scattering parameters. 
Thus confirmed consistency of the theoretical with experimental results
enables consistent calculation of $\mu({\bf k_{i}},T_{s})$.

Figure  \ref{recovfg2} shows the temperature dependence of total heat 
transfer in HAS from Xe/Cu(111) normalized to 
vertical component of the projectile incident
energy, $E_{zi}=E_{i}\cos^{2}\theta_{i}$. The $T_{s}$-dependence of 
$\mu_{rec}({\bf k_{i}},T_{s})$ hinders energy 
transfer to phonons and causes negative slopes in the plots.
This arises from larger phase space for projectile $c\rightarrow c$ transitions into
final states with $E_{z}>E_{zi}$. There it may give rise to negative $\mu({\bf
k_{i}},T_{s})$ (e.g. for $E_{i}=2.4$ meV and $\theta_{i}=50^{0}$ at $T_{s}>62$ K) and hence to heating of the scattered beam. In the classical theory this effect is 
independent of the accommodation coefficient and hence of $\theta_{i}$ 
\cite{Cerc}.
Here, the universal behaviour of $\mu({\bf k_{i}},T_{s})$ for higher 
$E_{i}$, as exemplified by the near 
coincidence of the two highest energy curves in Fig.  \ref{recovfg2} and 
confirmed by additional calculations at high $E_{i}$, manifests
itself only for {\em fixed} 
$\theta_{i}$ because the $\theta_{i}$-dependence of 
three-dimensional scattering matrix
elements is not contained solely in the factorizable 
scaling factor  $E_{zi}$ \cite{Xe,HAS}. Also, extension  of the 
classical Baule expression pertinent to energy transfer in the cubes
model ($\Delta{\bf K}=0$) to quantum surface
scattering\cite{CelliHimes} is generally inadequate, as is demonstrated
in comparison with our results for $T_{s}=0$.  
Since $U(z)$ with the well depth of 6.6 meV supports three bound states  we show in the inset the low-$E_{i}$ dependence of the recovery 
temperature (for which $\mu({\bf k_{i}},T_{r})=0$) calculated for the present phonon heatbath for
fixed $\theta_{i}$, for He coupling 
either only to S- or to S-, L-
and SH-phonon modes. The small difference indicates that  the major 
contribution is from strong He atom coupling to vertically polarized S-
\vskip 3.5 cm
\noindent
modes \cite{Xe}. Rapid variations in $T_{r}$ are caused by kinematic focusing in S-phonon assisted $c\rightarrow b$ transitions for large parallel momentum transfer.

\vskip 7.7 cm
\begin{figure}
\caption{Temperature dependence of the total energy transfer $\mu$ in He$\rightarrow$Xe/Cu(111) collisions
normalized to $E_{zi}$ for $\theta_{i}=50^{0}$ 
and  $E_{i}=80$ meV (full curve), 60 meV (long-dashed curve), 20 meV 
(short-dashed curve) and 2.4 meV (dash-dotted curve). 
The corresponding normalized values of $\mu$ at $T_{s}=0$ obtained from
the Baule formula corrected for the well depth are denoted by inverted triangle, circle, diamond and triangle, respectively. The scattered beam
is heated
on the average if $\mu<0$.
Inset: relative contributions of the phonon modes to
the recovery temperature $T_{r}$ (see text).}
\label{recovfg2}
\end{figure}

For monoatomic gases the standard classical accommodation theory gives 
$T_{r}=E_{i}/2k_{B}$ and to a good
approximation $E_{i}\simeq 5k_{B}T_{0}/2$ where $T_{0}$ is the stagnation 
temperature of beam gas prior to expansion in the nozzle
\cite{Legge}. This yields the recovery factor   
$T_{r}/T_{0}\simeq 1.25$ but deviations from this universal behaviour
have been observed in wind tunnel molecular beam experiments
\cite{Legge} and their explanation was proposed in terms of
heuristically modified classical expressions \cite{Cerc}. 
The present quantum theory enables essential progress beyond the
classical results by allowing the parallel momentum exchange with
phonons, multiphonon interference and quantum recoil of the projectile.
Their interplay gives the recovery factor as a function of $E_{i}$ 
(or $T_{0}$) and $\theta_{i}$ which for our prototype heatbath is shown in Fig. \ref{recovfg3}. Quite
generally, $T_{r}$ is largest for normal incidence and only at higher 
$E_{i}$ (i.e. $T_{0}$) quantum results may approach the classical limit
so far observed only for rough technical surfaces \cite{Legge}. Large deviations from the classical limit at low $E_{i}$ are due to
the quantum regime which allows larger $\Delta{\bf K}$ and transitions affected by the
bound states of He-surface potential.
Qualitatively these findings are not system specific as the present theory
is quite general and can be readily extended to calculations of the
heat transfer in collisions of He \cite{Siber,Wroclaw} or heavier rare
gas atoms \cite{commscat,Andersson} with clean surfaces in a wide range of 
scattering conditions.

\vskip 7 cm

\begin{figure}
\caption{Recovery factor $T_{r}/T_{0}$ for a prototype heatbath sustaining phonon modes typical of Xe/Cu(111) system plotted as a function of the incident energy 
$E_{i}$ or stagnation temperature $T_{0}$ of the gas for three
representative incident angles $\theta_{i}$. Dashed-dotted line is the 
classical result $T_{r}/T_{0}=1.25$.}
\label{recovfg3}
\end{figure}

In summary, we have shown that in inelastic gas-surface scattering
under the conditions of free
molecular flow the combination of quantum and temperature effects gives
rise to a violation of the universality of the energy transfer and
recovery factor predicted by the classical accommodation theory.

The work in Zagreb has been supported in part by the NSF grant JF-133.


\begin{references}


\bibitem{Hulpke} See the articles in: {\em Helium Atom Scattering from 
Surfaces}, Springer Series in Surface Sciences Vol. {\bf 27}, edited by 
E. Hulpke (Springer, Berlin, 1992).


\bibitem{Legge}H. Legge, J.P. Toennies and J. L\"{u}decke in 
{\it Rarefied Gas Dynamics 19 (1994)}, Vol.
II, edited by J. Harvey and G. Lord (Oxford Science Publications), 
p. 988; J.P. Toennies, {\it ibid}, p. 921; H. Legge, J.R. Manson and 
J.P. Toennies, J. Chem. Phys. {\bf 110},8767(1999).

\bibitem{Bird} G.A. Bird: {\em Molecular Gas Dynamics}, Claredon Press 
(Oxford, 1976).  


\bibitem{Schaaf} S.A. Schaaf and P.L. Chambr\'{e}: 
{\it Flow of Rarefied Gases}, Princeton University Press, Princeton, 
New Jersey, 1961. 

\bibitem{Cerc} C. Cercignani and M. Lampis, J. Appl. Math. Phys. 
{\bf 27},733(1976).


\bibitem{Hofmann} F. Hofmann, J.P. Toennies and J.R. Manson, J. Chem. 
Phys. {\bf 106},1234(1997). 

\bibitem{Sibener} K.D. Gibson and S.J. Sibener, Phys. Rev. Lett. 
{\bf 55},1514(1985).

\bibitem{GrahamJCP} 
C. Ramsayer. V. Pouthier, C. Girardet, P. Zeppenfeld, M. B\"{u}chel, 
V. Diercks and G. Comsa, Phys. Rev.
{\bf B55},13203(1997);
A.P. Graham, M.F. Bertino, F. Hofmann, J.P. Toennies and Ch. W\"{o}ll, 
J. Chem. Phys. {\bf 106},6194(1997).

\bibitem{BraunPRL}J. Braun, D. Fuhrmann, A. \v{S}iber, B. Gumhalter and
Ch. W\"{o}ll, Phys. Rev. Lett. {\bf 80},125(1998).

\bibitem{Xe} A. \v{S}iber, B. Gumhalter, J. Braun, A.P. Graham, M.F.
Bertino, J.P. Toennies, D. Fuhrmann and Ch. W\"{o}ll, Phys. Rev. 
{\bf B 59},5898(1999). 


\bibitem{BGL}  K. Burke, B. Gumhalter and D.C. Langreth, Phys. Rev. 
{\bf B 47},12852(1993).

\bibitem{HAS} A. Bili\'{c} and B. Gumhalter, Phys. Rev. 
{\bf B 52},12307(1995).

\bibitem{comment} B. Gumhalter and A. Bili\'{c}, Surf. Sci. 
{\bf 370},47(1997).

\bibitem{BB}J. B\"{o}heim and W. Brenig, Z. Phys. {\bf B 41},243(1981); T. Brunner and W. Brenig, Surf. Sci. {\bf 291},192(1993).

\bibitem{Boeheim} J. B\"{o}heim, Surf. Sci. {\bf 148},463(1984).

\bibitem{commscat} A. \v{S}iber and B. Gumhalter, Phys. Rev. Lett. 
{\bf 81},1742(1998).


\bibitem{dwf} B. Gumhalter, Surf. Sci. {\bf 347},237(1996).

\bibitem{AM} G. Armand and J.R. Manson, Phys. Rev. Lett. 
{\bf 53},1112(1984); ibid.
Surf. Sci. {\bf 195},513(1988). 

\bibitem{CelliHimes} V. Celli, D. Himes, P. Tran, J.P. Toennies, 
Ch. W\"{o}ll and G. Zhang, Phys. Rev. Lett. {\bf 66},3160(1991). 


\bibitem{Burke} C.A. DiRubio, D.M. Goodstein, B.H. Cooper and K. Burke,
Phys. Rev. Lett. {\bf 73},2768(1993).  



\bibitem{Siber} A. \v{S}iber and B. Gumhalter, Surf. Sci.  
{\bf 385},270(1997). 

\bibitem{Wroclaw} A. \v{S}iber, B. Gumhalter and J.P. Toennies, Vacuum 
{\bf 54}/315(1999).  

\bibitem{Andersson} F. Althoff, T. Andersson and S. Andersson, Phys.
Rev. Lett. {\bf 79},4429(1997).

\end{references}
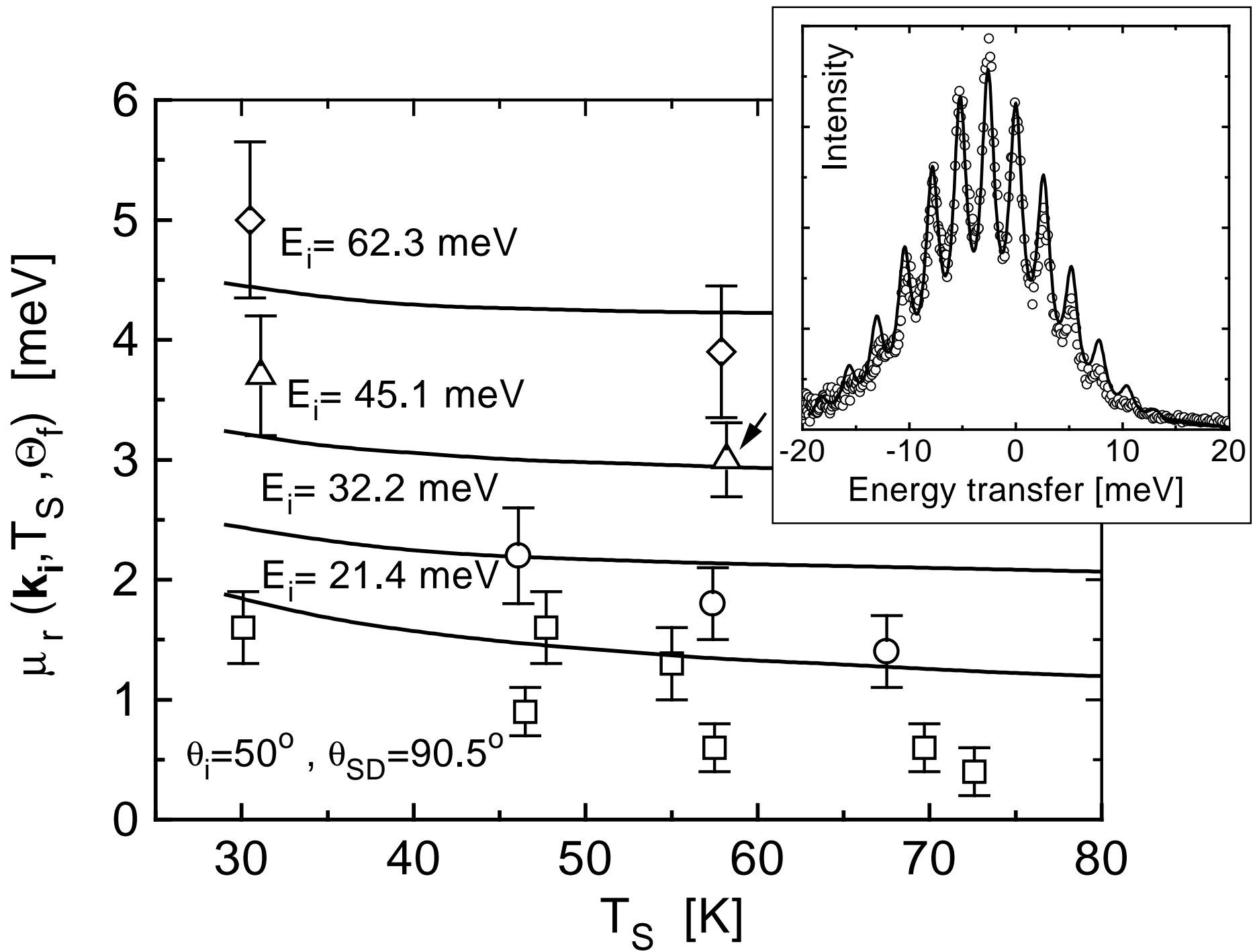
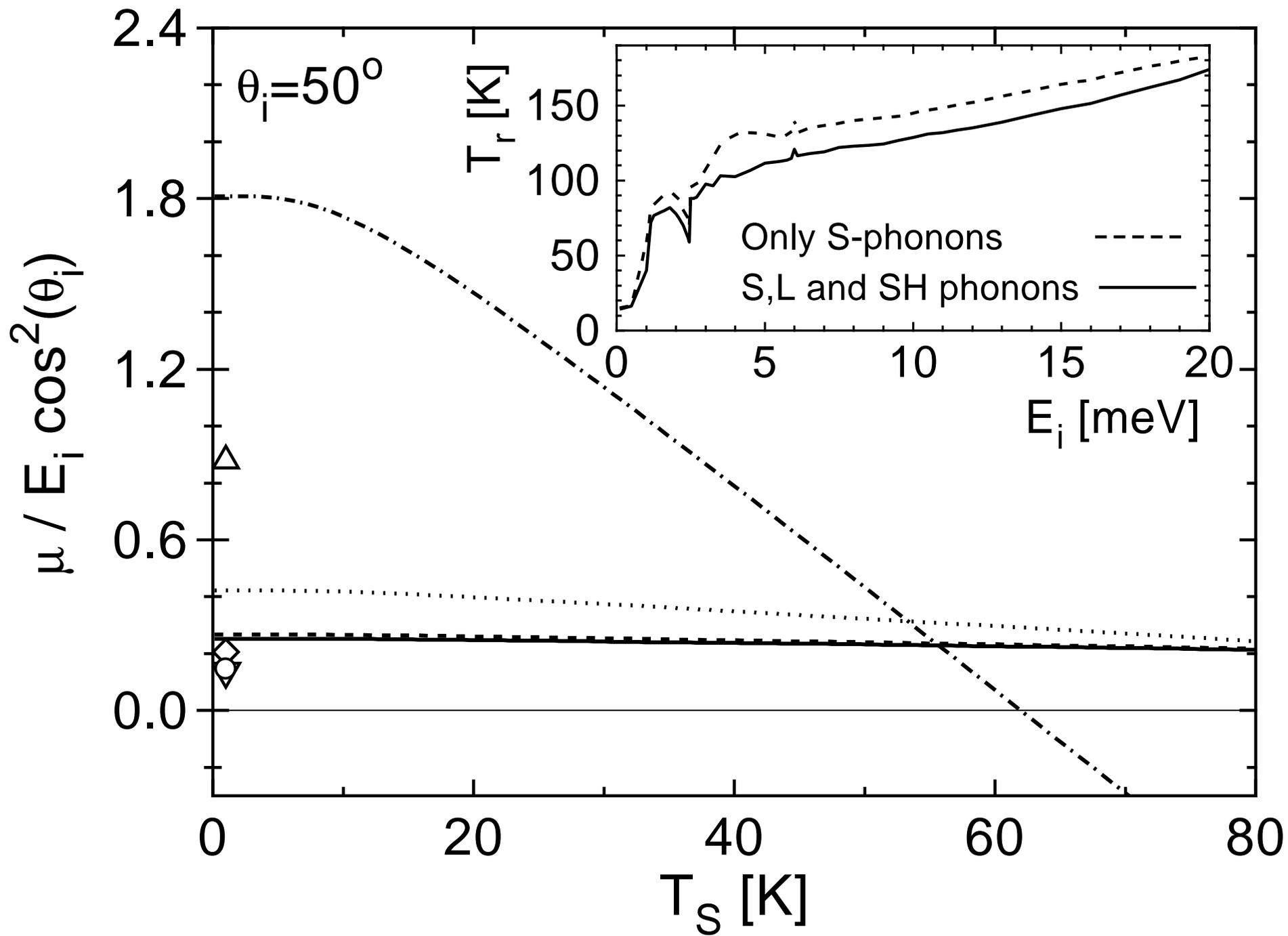
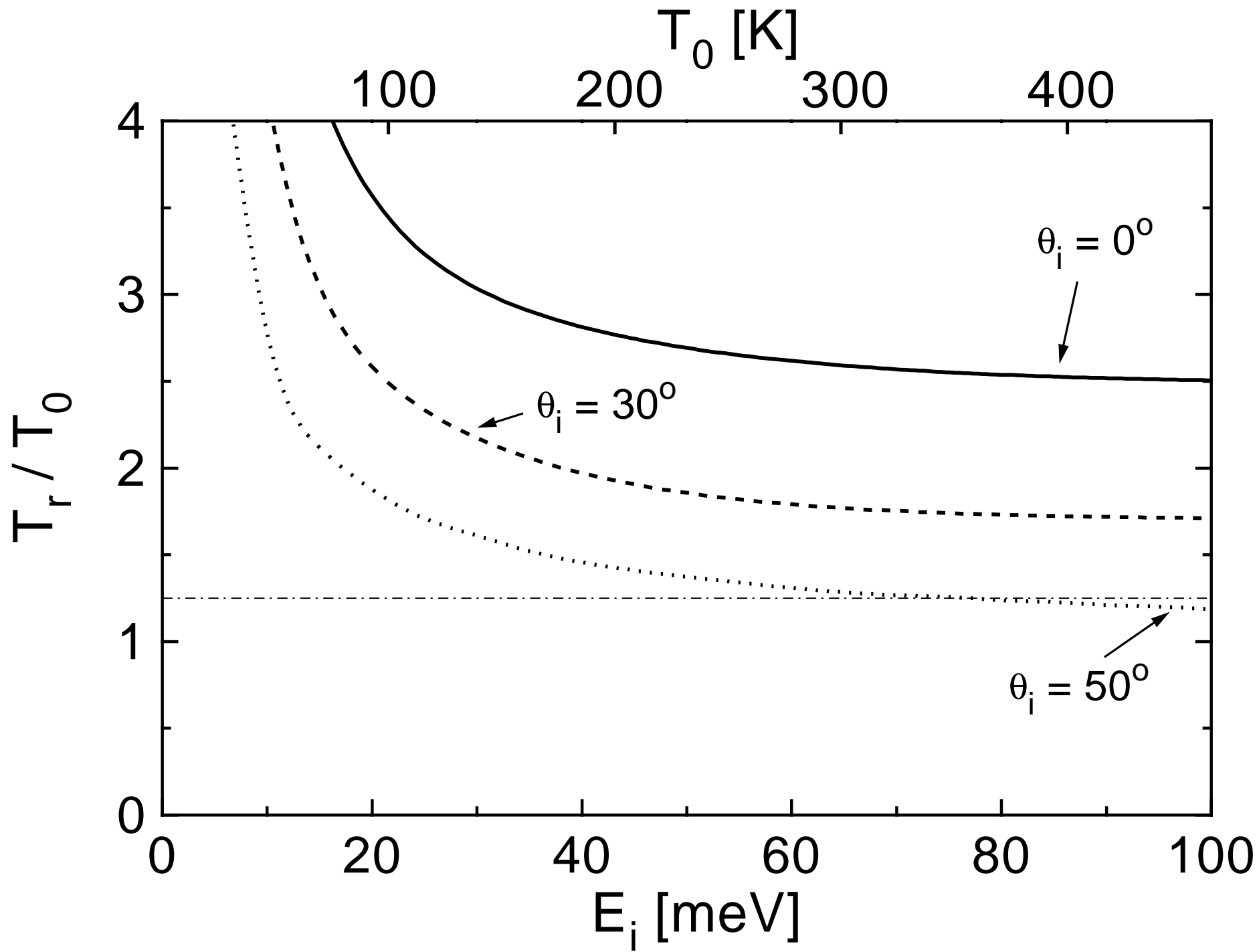
\end{document}